\documentclass[aps,prl,twocolumn,superscriptaddress,nofootinbib]{revtex4-1}
\usepackage[utf8]{inputenc}
\usepackage{amsmath}
\usepackage{amssymb}
\usepackage{epsfig}
\usepackage{graphicx,url}
\usepackage{bm}
\usepackage{mathrsfs}
\usepackage{color}
\usepackage[top=2cm, bottom=2.5cm, left=2cm, right=2cm]{geometry}
\usepackage{mathtools}
\usepackage{empheq}
\usepackage{physics}

\begin{document}
	
	\title{Non-Markovian out-of-equilibrium dynamics: A general numerical procedure to construct time-dependent memory kernels for coarse-grained observables}
	
	\author{Hugues Meyer}
	\affiliation{\it Physikalisches Institut, Albert-Ludwigs-Universit\"{a}t,  79104 Freiburg, Germany}
	\affiliation{Research Unit in Engineering Science, Universit\'{e} du Luxembourg,\\  L-4364 Esch-sur-Alzette, Luxembourg}
	\author{Philipp Pelagejcev}
	\affiliation{\it Physikalisches Institut, Albert-Ludwigs-Universit\"{a}t,  79104 Freiburg, Germany}
	\author{Tanja Schilling}
	\affiliation{\it Physikalisches Institut, Albert-Ludwigs-Universit\"{a}t,  79104 Freiburg, Germany}
	
	\date{\today}
	
	\begin{abstract}
	We present a numerical method to compute non-equilibrium memory kernels based on experimental data or molecular dynamics simulations. The procedure uses a recasting of the non-stationary generalized Langevin equation, in which we expand the memory kernel in a series that can be reconstructed iteratively. Each term in the series can be computed based solely on knowledge of the two-time auto-correlation function of the observable of interest. As a proof of principle, we apply the method to crystallization from a super-cooled Lennard Jones melt. We analyze the nucleation and growth dynamics of crystallites and observe that the memory kernel has a time extent that is about one order of magnitude larger than the typical timescale needed for a particle to be attached to the crystallite in the growth regime.
	\end{abstract}
	
	   \maketitle
	   
{\it Introduction} - The dynamics of complex many-body systems are often modelled in terms of the effective dynamics of a small set of relevant observables. Depending on the context, these observables are called 'reaction coordinates' or 'order parameters'\cite{Peters2016, Rohrdanz13, Sittel18}. E.g.~in biophysics one might be interested in the evolution of the shape of a protein during a folding experiment, but not in the motion of every single water molecule. Or in materials science, one might model the dynamics of a phase transition in terms of a suitable order parameter without resolving the details of the microscopic motion of the atoms. In the 1960's Mori and Zwanzig developed a projection operator formalism to derive the equation of motion of such averaged observables, the Generalized Langevin Equation (GLE) \cite{zwanzig:1961, mori:1965}. The GLE is valid only if the density of microstates is stationary. Based on the same formalism, but for non-stationary densities of microstates \cite{grabert:1982}, we have recently derived a general structure for the equation of motion of reaction coordinates \cite{meyer:2019,meyer:2017}. The structure applies to any dynamical process for which the microscopic equations of motion are deterministic and for any phase-space observable, even if it contains an explicit dependence on time. The resulting non-stationary Generalized Langevin Equation (nsGLE) is thus the equation that needs to be solved, if one wishes to derive reaction coordinate dynamics outside of thermal equilibrium. 

The GLE and the nsGLE contain an effective friction term that is non-local in time and integrates over the history of the process. The function that controls these history effects is called ``the memory kernel''. The non-locality in time makes the analysis of the GLE and the nsGLE mathematically inconvenient. In applications the memory kernel is therefore often approximated by a Dirac delta-distribution such that a time-local Langevin Equation is recovered \cite{huopaniemi:2006,horger:2008, espanol:2009, shea:2011, knoch:2015}. This assumption, that we refer to as the ``Markovian approximation'', is however in practice often not verified before it is used, which potentially leads to inaccurate or wrong results. To go beyond this approximation, there are two possible routes. The first one consists in searching (or constructing) a set of new reaction coordinates for which a Markovian description is correct \cite{peters:2006, rohrdanz:2011, perez:2013, peters:2016, chodera:2014}. This method is often useful, but the reaction coordinates constructed may either be abstract quantities, which are not guaranteed to be accessible in experiment, or a large set of them might be required, which implies that the Markovian Langevin equation needs to be solved in a high dimensional space \cite{Hegger09, Schaudinnus15, Schaudinnus16}. The second route consists in keeping the original observable despite its non-Markovian dynamics, and finding a way to evaluate the corresponding memory kernel. Various methods in this spirit do already exist, but they are restricted to processes with stationary microstate distributions, i.e.~they assume that the memory kernel is invariant under translations in time. We present here a general method to compute memory kernels for arbitrary processes and for arbitrary observables from MD simulations or experimental data, i.e.~a procedure to analyze the out-of-equilibrium dynamics of reaction coordinates and to construct their equation of motion.

The main idea of the procedure is to recast the nsGLE in a form in which the memory kernel is expressed as a sum of convolution products. Each term is determined recursively from the previous ones, and it probes the behaviour of memory at successively longer times. The discrepancies from the Markovian limit can be assessed by analyzing the functional shape of the computed terms and their relative timescales. The method neither requires modeling nor approximation and uses as a single input the two-time auto-correlation function of the observable under study, which is easily accessible in simulations or experiments. We also note that the computational time of the procedure scales linearly with the number of terms computed in the expansion.

As a proof of principle, we apply this new technique to the process of crystallization of a super-cooled Lennard-Jones melt. Crystallization processes are usually described in the framework of Transition State Theory, which is based on a Markovian approximation. We test this assumption here and reconstruct the memory kernel of the nsGLE which governs the evolution of nucleation and growth of crystalline clusters. We observe significant memory effects.

{\it Numerical method} - Consider a system of $N \gg 1 $ degrees of freedom $\left\{\Gamma_{i}\right\}_{i\leq N}$ that evolve according to deterministic microscopic equations of motion (e.g.~Hamilton's equations of motion), and a phase-space observable $A(\mathbf{\Gamma})$ that is fully determined by the location $\mathbf{\Gamma}$ in phase-space. Next, take an ensemble, i.e.~a large number of copies of the system, and allow these copies to be initialized according to any phase-space distribution, in particular a non-stationary one. The microscopic equations of motion as well as the observable $A$ can also be explicitly time-dependent. These formal definitions are general enough to encompass a very broad spectrum of processes. 
We showed in ref.~\cite{meyer:2019} that for any such observable and for any dynamical process, regardless of how far from equilibrium it evolves, one can always define functions $\omega(t)$, $K(t',t)$ and $\eta_{t}$ such that the equations of motion for $A_{t}$ and for its auto-correlation function $C(t',t) = \left\langle A^{*}(t') A(t) \right\rangle$ are 
\begin{align}
    \label{EOM_A}
    \frac{d A_{t}}{dt} &= \omega (t) A_{t} + \int_{0}^{t} d\tau  K(\tau, t) A_{\tau} + \eta_{t} \\
    \label{EOM_C}
    \frac{\partial C(t',t)}{\partial t} &= \omega (t) C(t',t) + \int_{t'}^{t} d\tau C(t',\tau) K(\tau, t)
\end{align}
where $K(\tau,t)$ is the memory kernel and the average is taken over the ensemble of non-equilibrium trajectories.
The time-dependence as a subscript denotes the dependence on a single trajectory, whereas the time-dependence between parentheses indicates a fixed function of time independent of the trajectory. In particular we showed $\omega(t) = \text{d}\left( \ln \sqrt{C(t,t)} \right)/\text{d}t$.  If the timescale on which $A_{t}$ evolves is much longer than the typical microscopic timescale, $\eta_{t}$ can be interpreted as a noise.

Several methods are already available to infer memory kernels from simulation data, e.g.~Fourier-Laplace analysis \cite{schnurr:1997,mokshin:2005, shin:2010, cordoba:2012}, projection operator analysis \cite{darve:2009, carof:2014, li:2017}, parametrization techniques \cite{lei:2016, ma:2016}, and iterative numerical inference schemes \cite{jung:2017}. However, these methods are applicable only to stationary processes where the memory kernel $K(t',t)$ effectively depends only on the difference $t'-t$. We propose here a method that applies in both the stationary and the non-stationary case. 
The basic idea is to use a measured (or simulated) auto-correlation function $C(t',t)$ to construct the memory kernel by means of an iterative procedure:

We formally integrate eqn.~(\ref{EOM_C}) into
\begin{equation}
    \label{EOM_C_J}
C(t',t) =  C(t',t') + \int_{t'}^{t}d\tau  C(t',\tau) \mathcal{J}(\tau, t)
\end{equation}
where we have defined
\begin{equation}
\mathcal{J}(t',t) = \omega(t') + \int_{t'}^{t} d\tau K(t', \tau)
\end{equation}
such that $K(t',t) = \partial_{t}\mathcal{J}(t',t)$. Taking the derivative  of eq.~(\ref{EOM_C_J}) with respect to $t'$ and rearranging terms allows to write
\begin{equation}
\label{J_first_iter}
\mathcal{J}(t',t) =  j_{0}(t',t) +  \int_{t'}^{t}d\tau \mathcal{S}_{0}(t',\tau) \mathcal{J}(\tau, t)  \\
\end{equation}
where we have defined
\begin{align}
\label{S0}
\mathcal{S}_{0}(t',t) &= C(t',t')^{-1} \partial_{t'} C(t',t) \\
\label{j0}
j_{0}(t',t) &=  C(t',t')^{-1} \partial_{t'}\left[ C(t',t') - C(t',t) \right]
\end{align}
(Note that the first time-derivative of the auto-correlation function $\partial_{t'} C(t',t)$ is all that is required for the following steps. We do not need to take any further time-derivatives of the input data.)

Next we substitute $\mathcal{J}(t,\tau)$ on the right-hand side of eq.~(\ref{J_first_iter}) by eq.~(\ref{J_first_iter}) itself.
\begin{align}
\mathcal{J}(t',t) =  j_{0}(t',t) &+ \int_{t'}^{t}d\tau \mathcal{S}_{0}(t',\tau) j_{0}(\tau,t)  \nonumber \\
&+ \int_{t'}^{t}d\tau \mathcal{S}_{1}(t',\tau) \mathcal{J}(\tau,t)
\end{align}
where we have defined $\mathcal{S}_{1}(t',t) =  \int_{t'}^{t}d\tau  \mathcal{S}_{0}(t',\tau) \mathcal{S}_{0}(\tau,t)$. This construction is similar to a typical diagrammatic expansion in liquid state theory, e.g.~in the context of the Ornstein-Zernike equation \cite{hansen:1990}. $\mathcal{S}_{1}(t',t)$ takes into account correlations at intermediate times $\tau$ similar to the way correlations between two particles in a liquid would be mediated by other particles.

We iterate the recursive substitution of $\mathcal{J}(\tau,t)$ below the integral and finally obtain
\begin{equation}
\label{J_full}
\mathcal{J}(t',t)   = j_{0}(t',t) +  \int_{t'}^{t}d\tau \mathcal{S}(t',\tau) j_{0}(\tau,t)
\end{equation}
where $\mathcal{S}(t',t)  := \sum_{n=0}^{\infty}  \mathcal{S}_{n}(t',t)$, and the functions $\mathcal{S}_{n}$ are defined recursively via the identity
\begin{equation}
\label{recursive_S}
\int_{t'}^{t}d\tau \mathcal{S}_{n}(t',\tau) \mathcal{S}_{m}(\tau,t) = \mathcal{S}_{n+m+1}(t',t),
\end{equation}
which is valid for any $(n,m)\in\mathbb{N}^{2}$. 
Note that in the limit $\omega = 0$ we have $\mathcal{J}(t',t)   \stackrel{\omega=0}{\to}  - \mathcal{S}(t',t)$. % In the Markovian limit, $K(t',t) = 2\gamma(t)\delta(t'-t)$. We show in the appendix that then the functions $\mathcal{S}_{n}$ are given by $\mathcal{S}_{n}(t',t) = (-1)^{n}\gamma(t') \int_{t'}^{t}d\tau_{1} \gamma (\tau_{1}) \cdots \int_{t'}^{\tau_{n-1}}d\tau_{n} \gamma(\tau_{n}) e^{\int_{t'}^{t} d\tau\gamma(\tau)}$, for $t>t'$, such that $\mathcal{S}(t',t) = -\gamma(t)$. In particular, assume $\gamma(t)$ is of constant sign for $t>t$ (which is in practice very often true) $\mathcal{S}_{0}(t',t)$ should be discontinuous at $t=t'$ and a monotonic function of $t$ for $t>t'$.
 In general, higher orders in the expansion have impact on longer times, and the number of terms needed for the sum $\mathcal{S}$ to converge at a certain time yields information about the strength and the time-extend of the memory effects. Details about this statement as well as about the Markovian limit are given in appendix.

As all terms on the right-hand side of eqn.~(\ref{J_full}) are expressed in terms of the auto-correlation function $C(t',t)$, we propose the following numerical scheme to compute $K(t',t)$:
\begin{enumerate}
\item Carry out a set of simulations (or experiments) and measure the observable $A_{t}$ for each trajectory. \footnote{The number of trajectories must be chosen such that the initial phase-space distribution $\rho_{0}(\mathbf{\Gamma})$ is probed with sufficient precision.}
\item Compute the two-time auto-correlation function $C(t',t) = N_{\text{traj}}^{-1} \sum_{ i\in \text{traj}} A_{t'}^{*(i)} A_{t}^{(i)}$.
\item Compute $\mathcal{S}_{0}$ and $j_{0}$ using eqns.~(\ref{S0}) and (\ref{j0}).
\item Compute  $\mathcal{S}_{n+1}$ recursively using eqn.~(\ref{recursive_S}) with $m=0$. Stop whenever for any times $t'$ and $t$, $\mathcal{S}_{n}(t',t)\ll \sum_{j=0}^{n}\mathcal{S}_{j}(t',t)$.
\item Compute $\mathcal{J}$ using eqn.~(\ref{J_full}) and finally $K(t',t) = \partial_{t}\mathcal{J}(t',t)$.
\end{enumerate}
This method is general enough to be applied to any dynamical process and any observable, even far from equilibrium, because it relies only on the structure of equs.~(\ref{EOM_A}) and (\ref{EOM_C}). Note that once the memory kernel has been constructed, the corresponding nsGLE can be solved to predict the dynamics of the process. One has thus obtained an effective coarse-grained description in terms of one coordinate rather than $N$.
	   
{\it Side Remark} - If one intends to study the fluctuations of $A_{t}$ independently from the evolution of the average trajectory, eqn.~(\ref{EOM_A}) may not be convenient. Instead one can use a modified, explictly time-dependent phase-space observable $\tilde{A}(\mathbf{\Gamma}, t)$ 
\begin{equation}
\label{modified_A}
\tilde{A}(\mathbf{\Gamma}, t) \equiv \left[A(\mathbf{\Gamma}) - \mu_{A}(t)\right]/\sigma_{A}(t)
\end{equation}
where $\mu_{A}(t) =  \left\langle A(t) \right\rangle$ is the time-dependent average of $A$ and $\sigma_{A}(t)^{2} = \left\langle \left|A(t)\right|^{2} \right\rangle -  \left\langle A(t) \right\rangle\left\langle A^{*}(t) \right\rangle$ is its time-dependent variance. The obervable $\tilde{A}$ measure deviations from the average trajectory, normalized by the variance of the process, it is thus a unitless number indicating whether a particular trajectory is delayed or advanced compared to the average one. We can define the corresponding auto-correlation function as $\tilde{C}(t',t) = \left\langle \tilde{A}^{*}(t')\tilde{A}(t) \right\rangle$ (the notation $\tilde{\cdot}$  from this point on refers to any function related to $\tilde{A}$), which characterizes how fluctuations decorrelate irrespective of the amplitude of the variable $A$. The structure of both equ.~(\ref{EOM_A}) and equ.~(\ref{EOM_C}) remains valid for $\tilde{A}$ and $\tilde{C}$. We thus introduce a new memory kernel $\tilde{K}$, a fluctuating force $\tilde{\eta}_{t}$ with $\left\langle \tilde{\eta}(t) \right\rangle = 0$, and a drift function $\tilde{\omega}$ which turns out to vanish because $\tilde{C}(t,t) = 1$ is constant, yielding
\begin{equation}
\label{EOM_A_tilde}
\frac{d \tilde{A}_{t}}{dt} = \int_{0}^{t} d\tau \tilde{K}(\tau, t) \tilde{A}_{\tau} + \tilde{\eta}_{t}
\end{equation}
This is then used to rewrite the equation of motion for $A_{t}$. Using $\sigma_{A}(t) d\tilde{A}_{t}/dt = d\Delta A_{t}/dt - \Delta A_{t}\dot{\sigma_{A}}(t)/\sigma_{A}(t)$, where $\Delta A_{t} = A_{t}-\mu_{A}(t)$, we can rewrite eqn.~(\ref{EOM_A_tilde})  
\begin{align}
\label{EOM_A_2}
\frac{dA_{t}}{dt} =& \dot{\mu}_{A}(t) + \Omega(t) \Delta A_{t} + \int_{t'}^{t} d\tau \Delta A_{\tau}  \mathcal{K}(\tau,t) + f_{t'}(t',t)
\end{align}
with $\Omega(t) = d \ln\left(\sigma_{A}(t) \right)/dt$, $\mathcal{K}(t',t) = \frac{\sigma_{A}(t)}{\sigma_{A}(t')} \tilde{K}(t',t)$ and $f_{t'}(t',t) = \sigma_{A}(t) \tilde{\eta}_{t'}(t',t)$.
Hence there is a direct correspondence between the equation of motion for $\tilde{A}$ and the one for $A$. Note that $\omega\neq \Omega$ and $K \neq \mathcal{K}$ such that eqns.~(\ref{EOM_A}) and (\ref{EOM_A_2}) are not identical, but their mathematical structure is the same. When studying memory effects and deriving an equation of motion for an observable, one can chose between these equivalent descriptions. In both cases, the method to numerically infer memory kernels is valid.

\begin{figure}
\includegraphics[width=\linewidth]{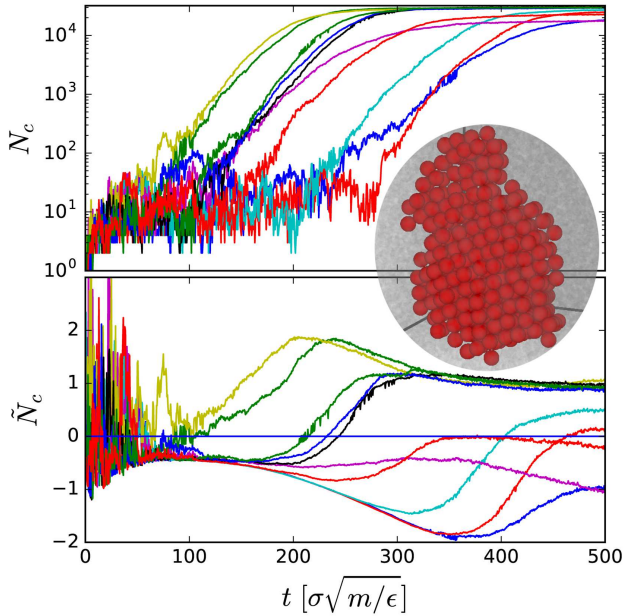}
\caption{Evolution of the size of the largest cluster $N_{c}$ (top), and its associated modified observable $\tilde{N}_{c}$, for several simulation trajectories. We also show an example snapshot of a crystallite surrounded by the liquid phase.}
\label{trajectories}
\end{figure}

{\it Lennard-Jones crystallization} - As a test of the numerical procedure we analyzed the dynamics of crystallization.
We carried out molecular dynamics simulations of $N=32,000$ particles interacting via a 6-12 Lennard-Jones potential. We used a cubic box with periodic boundary conditions and we ran the dynamics in the NVT ensemble, using a Nosé-Hoover thermostat to control the temperature. We first equilibrated the liquid phase at density $\rho = 1$ and temperature $T=2$ (in Lennard-Jones reduced units), and then instantaneously quenched the temperature to $T=0.75$, for which the equilibrium phase is known to be a crystal. We then let the system evolve freely (i.e.~we did not use any biasing scheme to speed up sampling). In order to monitor the formation and growth of crystallites, we used the method introduced by ten Wolde et.\ al \cite{tenwolde:1996} based on orientational bond order parameters \cite{steinhardt:1983}. A crystalline cluster is defined as a set of neighbouring crystal-like particles. The observable that we used as a reaction coordinate is the number $N_{c}$ of particles in the largest crystalline cluster. We observed the evolution of $N_{c}(t)$, see fig.~\ref{trajectories}, upper panel, where we also show a typical snapshot of a crystallite. A total of 4019 trajectories were used for the analysis.  

\begin{figure}
\begin{center}
  \includegraphics[width=\linewidth]{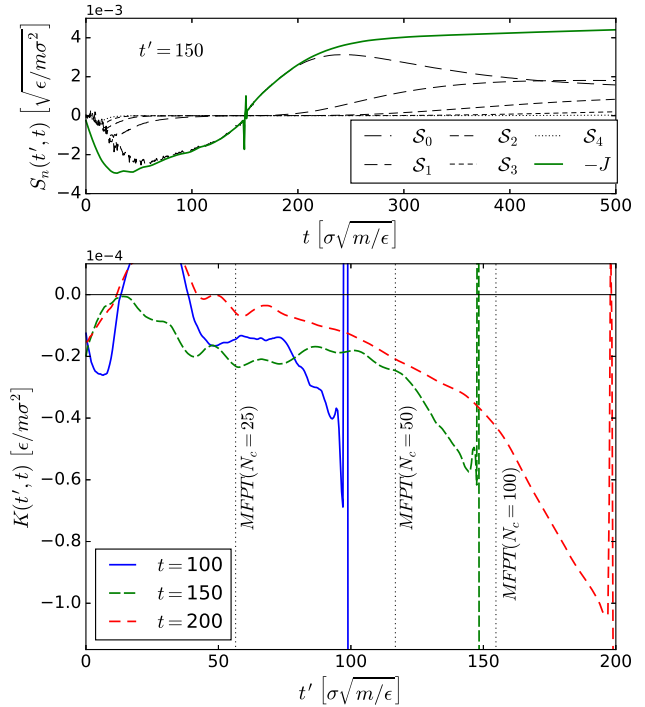}
  \caption{\underline{Top} : $\mathcal{S}_{n}(t',t)$ for $0\leq n \leq 4$ as well of their sum $\mathcal{S}(t',t) = \sum_{n}\mathcal{S}_{n}(t',t) = -J(t',t)$, as a function of $t$ for a fixed $t'=150$.  $\mathcal{J}$ would be a step function of $t$ in the Markovian case.  \underline{Bottom} : Memory kernel $K(t',t)$ as a function of $t'$. We also show the values of the mean first passage times for different values of $N_{c}$ as vertical lines. According to committor analysis  the critical nucleus has $N_{c} \approx 80$, thus $t=100$ is in the induction time regime and $t=200$ in the growth regime.  % Note also that from eqn.~(\ref{EOM_A}), the term memory is meaningful only if we look at $t'<t$, both for $\mathcal{J}$ and $K$.
  }
\label{kernels}
\end{center}
\end{figure}

We computed for each trajectory the modified variable $\tilde{N}_{c}$ as defined in eqn.~(\ref{modified_A}) (see fig.~\ref{trajectories}, lower panel), and its auto-correlation function $\tilde{C}(t',t)$. We then applied the numerical method presented above. As we used $\tilde{C}(t',t)$, the drift term $\tilde{\omega}$ vanishes, which implies $j_{0} = -\mathcal{S}_{0}$ and $\mathcal{J}= -\mathcal{S}$. We computed $S_{0}(t',t) = \partial_{t'}\tilde{C}(t',t)$, and we then iteratively applied the recursion relation eq.~(\ref{recursive_S}) with $m=0$, i.e. $\int_{t'}^{t}d\tau \mathcal{S}_{0}(t',\tau) \mathcal{S}_{n}(\tau,t) = \mathcal{S}_{n+1}(t',t)$. We show in fig.~(\ref{kernels}) the functions $\mathcal{S}_{n}$, as well as the function $\mathcal{J}= -\sum_{n} S_{n}$. If the nucleation process were Markovian, $\mathcal{J}$ would be a step function of $t$, constant for $t<t'$ and $t>t'$ and discontinuous at $t=t'$. Although we observe this discontinuity, the behaviour around $t\neq t'$ is far from being constant. In addition, we note that $\tilde{N}_{c}(t)$ vanishes for each trajectory in the limit $t\to 0$, which implies that $\mathcal{S}_{0}(t',t) \sim 0$ for $t\to 0$, and thus $J(t',t) = -\mathcal{S}(t',t) \sim 0$ for $t \to 0$. This is observed numerically, and suggests a non-trivial behaviour for the memory kernel $K$ at short times. The algorithm needed 15 iterations in $\mathcal{S}$ to properly converge.  
The functions $S_{n}(t',t)$ probe the behaviour of memory effects for different times. Each new order $n$ presents extrema at increasing values of $|t'-t|$. The summation of all the orders results in the smooth function $\mathcal{S}=-\mathcal{J}$ (see fig.~\ref{kernels}). % (Note that the singularity at $t'=t$ implies slight numerical inacurracies.)

\begin{figure}
\begin{center}
\includegraphics[width=\linewidth]{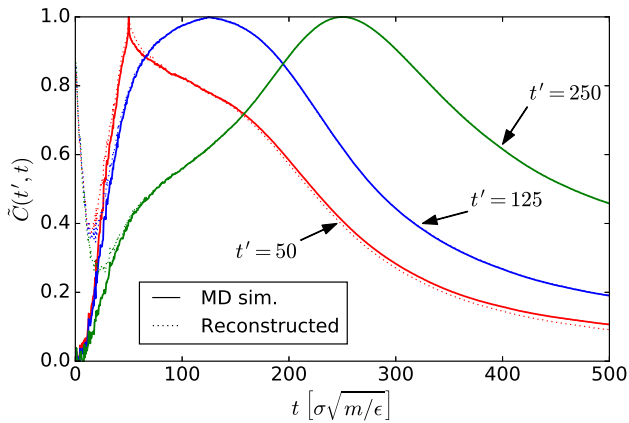}
\caption{Auto-correlation function $C(t',t)$ for various values of $t'$ as a function of $t$. The solid lines are directly computed from the MD simulations, the dotted lines are the right-hand side of eqn.~(\ref{EOM_C_J}), where $\mathcal{J}$ is computed via the method presented. The overlap is nearly perfect. 
}
\label{reconstruction}
\end{center}
\end{figure}

We checked the validity of the function $\mathcal{J}$ obtained in this way by using it as an input to compute the right-hand side of eqn.~(\ref{EOM_C_J}), and we compared the result to the left-hand side, i.e. $\tilde{C}(t',t)$ itself. As shown in figure~\ref{reconstruction}, the overlap is very good \footnote{The discrepancies at very short times are due to numerical errors in the computation of the derivative of $\tilde{C}(t',t)$. Due to the initial vanishing width of the crystllite size distribution, the number of trajectories needed to eliminate statistical fluctuations at short times is very large. Here our aim is to demonstrate the validity of our method and not capture the details of the nucleation process, therefore we did not run additional simulations to resolve the early stage dynamics.}. This test confirms that the method presented is able to reconstruct the dynamics of a reaction coordinate, and that it can be used to develop numerical coarse-graining schemes for dynamics out of thermal equilibrium. 


Once this check on $\mathcal{J}(t',t)$ was performed, we computed its derivative with respect to $t$, yielding $K(t',t)$ which is also shown in fig.~(\ref{kernels}). In addition we show vertical lines indicating the mean first passage times for various values of $N_{c}$. A committor analysis of the trajectories showed that the critical crystallite contains ca.~80 particles. Hence we observe memory both, in the nucleation and in the growth regime. % We can also compare the extend of the memory kernel to the average time needed for a particle to be attached to the droplet in the growth regime.
% We computed, for each trajectory an estimate of $\tau_{att} = \left\langle \dot{N_{c}}/S \right\rangle \propto \left\langle dN_{c}^{1/3}(t) /dt \right\rangle$ for a spherical droplet (where $S$ is the surface of the droplet).
The average time needed for a particle to be attached to a cluster surface area of $\sigma_{LJ}^2$ was $\tau_{att}\simeq 5 $ (in LJ units), which is about one order of magnitude smaller that the time extent of the memory kernel. We thus conclude that memory effects are not negligible in the Lennard-Jones crystallization process.

{\it Conclusion} - We have introduced a numerical method to construct memory kernels for any process for which the non stationary Generalized Langevin Equation is relevant, regardless of how far from equilibrium the system is. This procedure requires little computational effort and no modeling assumption since its only input is the two-time auto-correlation function of the observable under study. The method can also be applied to a modified version of the nsGLE, where the memory kernel and the fluctuating force contribute only to the fluctuations about the average of the observable of interest. We investigated the crystallization process as a proof of principle, and we have shown that the procedure allows to reconstruct the dynamics of the problem. In this particular example, we demonsrate that memory effects play a significant role for nucleation and growth dynamics.

{\it Acknowledgement} - We thank G.~Stock for valuable feedback on the manuscript.


%\bibliographystyle{unsrt}
%\bibliography{references3}

\begin{thebibliography}{34}%
\makeatletter
\providecommand \@ifxundefined [1]{%
 \@ifx{#1\undefined}
}%
\providecommand \@ifnum [1]{%
 \ifnum #1\expandafter \@firstoftwo
 \else \expandafter \@secondoftwo
 \fi
}%
\providecommand \@ifx [1]{%
 \ifx #1\expandafter \@firstoftwo
 \else \expandafter \@secondoftwo
 \fi
}%
\providecommand \natexlab [1]{#1}%
\providecommand \enquote  [1]{``#1''}%
\providecommand \bibnamefont  [1]{#1}%
\providecommand \bibfnamefont [1]{#1}%
\providecommand \citenamefont [1]{#1}%
\providecommand \href@noop [0]{\@secondoftwo}%
\providecommand \href [0]{\begingroup \@sanitize@url \@href}%
\providecommand \@href[1]{\@@startlink{#1}\@@href}%
\providecommand \@@href[1]{\endgroup#1\@@endlink}%
\providecommand \@sanitize@url [0]{\catcode `\\12\catcode `\$12\catcode
  `\&12\catcode `\#12\catcode `\^12\catcode `\_12\catcode `\%12\relax}%
\providecommand \@@startlink[1]{}%
\providecommand \@@endlink[0]{}%
\providecommand \url  [0]{\begingroup\@sanitize@url \@url }%
\providecommand \@url [1]{\endgroup\@href {#1}{\urlprefix }}%
\providecommand \urlprefix  [0]{URL }%
\providecommand \Eprint [0]{\href }%
\providecommand \doibase [0]{http://dx.doi.org/}%
\providecommand \selectlanguage [0]{\@gobble}%
\providecommand \bibinfo  [0]{\@secondoftwo}%
\providecommand \bibfield  [0]{\@secondoftwo}%
\providecommand \translation [1]{[#1]}%
\providecommand \BibitemOpen [0]{}%
\providecommand \bibitemStop [0]{}%
\providecommand \bibitemNoStop [0]{.\EOS\space}%
\providecommand \EOS [0]{\spacefactor3000\relax}%
\providecommand \BibitemShut  [1]{\csname bibitem#1\endcsname}%
\let\auto@bib@innerbib\@empty
%</preamble>
\bibitem [{\citenamefont {Peters}(2016{\natexlab{a}})}]{Peters2016}%
  \BibitemOpen
  \bibfield  {author} {\bibinfo {author} {\bibfnamefont {B.}~\bibnamefont
  {Peters}},\ }\href@noop {} {\bibfield  {journal} {\bibinfo  {journal} {Annual
  review of physical chemistry}\ }\textbf {\bibinfo {volume} {67}},\ \bibinfo
  {pages} {669} (\bibinfo {year} {2016}{\natexlab{a}})}\BibitemShut {NoStop}%
\bibitem [{\citenamefont {Rohrdanz}\ \emph {et~al.}(2013)\citenamefont
  {Rohrdanz}, \citenamefont {Zheng},\ and\ \citenamefont
  {Clementi}}]{Rohrdanz13}%
  \BibitemOpen
  \bibfield  {author} {\bibinfo {author} {\bibfnamefont {M.~A.}\ \bibnamefont
  {Rohrdanz}}, \bibinfo {author} {\bibfnamefont {W.}~\bibnamefont {Zheng}}, \
  and\ \bibinfo {author} {\bibfnamefont {C.}~\bibnamefont {Clementi}},\
  }\href@noop {} {\bibfield  {journal} {\bibinfo  {journal} {Annu. Rev. Phys.
  Chem.}\ }\textbf {\bibinfo {volume} {64}},\ \bibinfo {pages} {295} (\bibinfo
  {year} {2013})}\BibitemShut {NoStop}%
\bibitem [{\citenamefont {Sittel}\ and\ \citenamefont
  {Stock}(2018)}]{Sittel18}%
  \BibitemOpen
  \bibfield  {author} {\bibinfo {author} {\bibfnamefont {F.}~\bibnamefont
  {Sittel}}\ and\ \bibinfo {author} {\bibfnamefont {G.}~\bibnamefont {Stock}},\
  }\href@noop {} {\bibfield  {journal} {\bibinfo  {journal} {J. Chem. Phys.}\
  }\textbf {\bibinfo {volume} {149}},\ \bibinfo {pages} {150901} (\bibinfo
  {year} {2018})}\BibitemShut {NoStop}%
\bibitem [{\citenamefont {Zwanzig}(1961)}]{zwanzig:1961}%
  \BibitemOpen
  \bibfield  {author} {\bibinfo {author} {\bibfnamefont {R.}~\bibnamefont
  {Zwanzig}},\ }\href@noop {} {\bibfield  {journal} {\bibinfo  {journal}
  {Physical Review}\ }\textbf {\bibinfo {volume} {124}},\ \bibinfo {pages}
  {983} (\bibinfo {year} {1961})}\BibitemShut {NoStop}%
\bibitem [{\citenamefont {Mori}(1965)}]{mori:1965}%
  \BibitemOpen
  \bibfield  {author} {\bibinfo {author} {\bibfnamefont {H.}~\bibnamefont
  {Mori}},\ }\href@noop {} {\bibfield  {journal} {\bibinfo  {journal} {Progress
  of theoretical physics}\ }\textbf {\bibinfo {volume} {33}},\ \bibinfo {pages}
  {423} (\bibinfo {year} {1965})}\BibitemShut {NoStop}%
\bibitem [{\citenamefont {Grabert}(1982)}]{grabert:1982}%
  \BibitemOpen
  \bibfield  {author} {\bibinfo {author} {\bibfnamefont {H.}~\bibnamefont
  {Grabert}},\ }\href@noop {} {\emph {\bibinfo {title} {Projection operator
  techniques in nonequilibrium statistical mechanics}}},\ Vol.~\bibinfo
  {volume} {95}\ (\bibinfo  {publisher} {Springer},\ \bibinfo {year}
  {1982})\BibitemShut {NoStop}%
\bibitem [{\citenamefont {Meyer}\ \emph {et~al.}(2019)\citenamefont {Meyer},
  \citenamefont {Voigtmann},\ and\ \citenamefont {Schilling}}]{meyer:2019}%
  \BibitemOpen
  \bibfield  {author} {\bibinfo {author} {\bibfnamefont {H.}~\bibnamefont
  {Meyer}}, \bibinfo {author} {\bibfnamefont {T.}~\bibnamefont {Voigtmann}}, \
  and\ \bibinfo {author} {\bibfnamefont {T.}~\bibnamefont {Schilling}},\
  }\href@noop {} {\bibfield  {journal} {\bibinfo  {journal} {The Journal of
  chemical physics}\ }\textbf {\bibinfo {volume} {150}},\ \bibinfo {pages}
  {174118} (\bibinfo {year} {2019})}\BibitemShut {NoStop}%
\bibitem [{\citenamefont {Meyer}\ \emph {et~al.}(2017)\citenamefont {Meyer},
  \citenamefont {Voigtmann},\ and\ \citenamefont {Schilling}}]{meyer:2017}%
  \BibitemOpen
  \bibfield  {author} {\bibinfo {author} {\bibfnamefont {H.}~\bibnamefont
  {Meyer}}, \bibinfo {author} {\bibfnamefont {T.}~\bibnamefont {Voigtmann}}, \
  and\ \bibinfo {author} {\bibfnamefont {T.}~\bibnamefont {Schilling}},\
  }\href@noop {} {\bibfield  {journal} {\bibinfo  {journal} {The Journal of
  chemical physics}\ }\textbf {\bibinfo {volume} {147}},\ \bibinfo {pages}
  {214110} (\bibinfo {year} {2017})}\BibitemShut {NoStop}%
\bibitem [{\citenamefont {Huopaniemi}\ \emph {et~al.}(2006)\citenamefont
  {Huopaniemi}, \citenamefont {Luo}, \citenamefont {Ala-Nissila},\ and\
  \citenamefont {Ying}}]{huopaniemi:2006}%
  \BibitemOpen
  \bibfield  {author} {\bibinfo {author} {\bibfnamefont {I.}~\bibnamefont
  {Huopaniemi}}, \bibinfo {author} {\bibfnamefont {K.}~\bibnamefont {Luo}},
  \bibinfo {author} {\bibfnamefont {T.}~\bibnamefont {Ala-Nissila}}, \ and\
  \bibinfo {author} {\bibfnamefont {S.-C.}\ \bibnamefont {Ying}},\ }\href@noop
  {} {\bibfield  {journal} {\bibinfo  {journal} {The Journal of chemical
  physics}\ }\textbf {\bibinfo {volume} {125}},\ \bibinfo {pages} {124901}
  (\bibinfo {year} {2006})}\BibitemShut {NoStop}%
\bibitem [{\citenamefont {H{\"o}rger}\ \emph {et~al.}(2008)\citenamefont
  {H{\"o}rger}, \citenamefont {Velasco}, \citenamefont {Mingorance},
  \citenamefont {Rivas}, \citenamefont {Tarazona},\ and\ \citenamefont
  {V{\'e}lez}}]{horger:2008}%
  \BibitemOpen
  \bibfield  {author} {\bibinfo {author} {\bibfnamefont {I.}~\bibnamefont
  {H{\"o}rger}}, \bibinfo {author} {\bibfnamefont {E.}~\bibnamefont {Velasco}},
  \bibinfo {author} {\bibfnamefont {J.}~\bibnamefont {Mingorance}}, \bibinfo
  {author} {\bibfnamefont {G.}~\bibnamefont {Rivas}}, \bibinfo {author}
  {\bibfnamefont {P.}~\bibnamefont {Tarazona}}, \ and\ \bibinfo {author}
  {\bibfnamefont {M.}~\bibnamefont {V{\'e}lez}},\ }\href@noop {} {\bibfield
  {journal} {\bibinfo  {journal} {Physical Review E}\ }\textbf {\bibinfo
  {volume} {77}},\ \bibinfo {pages} {011902} (\bibinfo {year}
  {2008})}\BibitemShut {NoStop}%
\bibitem [{\citenamefont {Espanol}\ and\ \citenamefont
  {L{\"o}wen}(2009)}]{espanol:2009}%
  \BibitemOpen
  \bibfield  {author} {\bibinfo {author} {\bibfnamefont {P.}~\bibnamefont
  {Espanol}}\ and\ \bibinfo {author} {\bibfnamefont {H.}~\bibnamefont
  {L{\"o}wen}},\ }\href@noop {} {\bibfield  {journal} {\bibinfo  {journal} {The
  Journal of chemical physics}\ }\textbf {\bibinfo {volume} {131}},\ \bibinfo
  {pages} {244101} (\bibinfo {year} {2009})}\BibitemShut {NoStop}%
\bibitem [{\citenamefont {Shea}\ and\ \citenamefont
  {Kreuzer}(2011)}]{shea:2011}%
  \BibitemOpen
  \bibfield  {author} {\bibinfo {author} {\bibfnamefont {P.}~\bibnamefont
  {Shea}}\ and\ \bibinfo {author} {\bibfnamefont {H.~J.}\ \bibnamefont
  {Kreuzer}},\ }\href@noop {} {\bibfield  {journal} {\bibinfo  {journal}
  {Surface Science}\ }\textbf {\bibinfo {volume} {605}},\ \bibinfo {pages}
  {296} (\bibinfo {year} {2011})}\BibitemShut {NoStop}%
\bibitem [{\citenamefont {Knoch}\ and\ \citenamefont
  {Speck}(2015)}]{knoch:2015}%
  \BibitemOpen
  \bibfield  {author} {\bibinfo {author} {\bibfnamefont {F.}~\bibnamefont
  {Knoch}}\ and\ \bibinfo {author} {\bibfnamefont {T.}~\bibnamefont {Speck}},\
  }\href@noop {} {\bibfield  {journal} {\bibinfo  {journal} {New Journal of
  Physics}\ }\textbf {\bibinfo {volume} {17}},\ \bibinfo {pages} {115004}
  (\bibinfo {year} {2015})}\BibitemShut {NoStop}%
\bibitem [{\citenamefont {Peters}\ and\ \citenamefont
  {Trout}(2006)}]{peters:2006}%
  \BibitemOpen
  \bibfield  {author} {\bibinfo {author} {\bibfnamefont {B.}~\bibnamefont
  {Peters}}\ and\ \bibinfo {author} {\bibfnamefont {B.~L.}\ \bibnamefont
  {Trout}},\ }\href@noop {} {\bibfield  {journal} {\bibinfo  {journal} {The
  Journal of chemical physics}\ }\textbf {\bibinfo {volume} {125}},\ \bibinfo
  {pages} {054108} (\bibinfo {year} {2006})}\BibitemShut {NoStop}%
\bibitem [{\citenamefont {Rohrdanz}\ \emph {et~al.}(2011)\citenamefont
  {Rohrdanz}, \citenamefont {Zheng}, \citenamefont {Maggioni},\ and\
  \citenamefont {Clementi}}]{rohrdanz:2011}%
  \BibitemOpen
  \bibfield  {author} {\bibinfo {author} {\bibfnamefont {M.~A.}\ \bibnamefont
  {Rohrdanz}}, \bibinfo {author} {\bibfnamefont {W.}~\bibnamefont {Zheng}},
  \bibinfo {author} {\bibfnamefont {M.}~\bibnamefont {Maggioni}}, \ and\
  \bibinfo {author} {\bibfnamefont {C.}~\bibnamefont {Clementi}},\ }\href@noop
  {} {\bibfield  {journal} {\bibinfo  {journal} {The Journal of chemical
  physics}\ }\textbf {\bibinfo {volume} {134}},\ \bibinfo {pages} {03B624}
  (\bibinfo {year} {2011})}\BibitemShut {NoStop}%
\bibitem [{\citenamefont {P{\'e}rez-Hern{\'a}ndez}\ \emph
  {et~al.}(2013)\citenamefont {P{\'e}rez-Hern{\'a}ndez}, \citenamefont {Paul},
  \citenamefont {Giorgino}, \citenamefont {De~Fabritiis},\ and\ \citenamefont
  {No{\'e}}}]{perez:2013}%
  \BibitemOpen
  \bibfield  {author} {\bibinfo {author} {\bibfnamefont {G.}~\bibnamefont
  {P{\'e}rez-Hern{\'a}ndez}}, \bibinfo {author} {\bibfnamefont
  {F.}~\bibnamefont {Paul}}, \bibinfo {author} {\bibfnamefont {T.}~\bibnamefont
  {Giorgino}}, \bibinfo {author} {\bibfnamefont {G.}~\bibnamefont
  {De~Fabritiis}}, \ and\ \bibinfo {author} {\bibfnamefont {F.}~\bibnamefont
  {No{\'e}}},\ }\href@noop {} {\bibfield  {journal} {\bibinfo  {journal} {The
  Journal of chemical physics}\ }\textbf {\bibinfo {volume} {139}},\ \bibinfo
  {pages} {07B604\_1} (\bibinfo {year} {2013})}\BibitemShut {NoStop}%
\bibitem [{\citenamefont {Peters}(2016{\natexlab{b}})}]{peters:2016}%
  \BibitemOpen
  \bibfield  {author} {\bibinfo {author} {\bibfnamefont {B.}~\bibnamefont
  {Peters}},\ }\href@noop {} {\bibfield  {journal} {\bibinfo  {journal} {Annual
  review of physical chemistry}\ }\textbf {\bibinfo {volume} {67}},\ \bibinfo
  {pages} {669} (\bibinfo {year} {2016}{\natexlab{b}})}\BibitemShut {NoStop}%
\bibitem [{\citenamefont {Chodera}\ and\ \citenamefont
  {No{\'e}}(2014)}]{chodera:2014}%
  \BibitemOpen
  \bibfield  {author} {\bibinfo {author} {\bibfnamefont {J.~D.}\ \bibnamefont
  {Chodera}}\ and\ \bibinfo {author} {\bibfnamefont {F.}~\bibnamefont
  {No{\'e}}},\ }\href@noop {} {\bibfield  {journal} {\bibinfo  {journal}
  {Current opinion in structural biology}\ }\textbf {\bibinfo {volume} {25}},\
  \bibinfo {pages} {135} (\bibinfo {year} {2014})}\BibitemShut {NoStop}%
\bibitem [{\citenamefont {Hegger}\ and\ \citenamefont
  {Stock}(2009)}]{Hegger09}%
  \BibitemOpen
  \bibfield  {author} {\bibinfo {author} {\bibfnamefont {R.}~\bibnamefont
  {Hegger}}\ and\ \bibinfo {author} {\bibfnamefont {G.}~\bibnamefont {Stock}},\
  }\href@noop {} {\bibfield  {journal} {\bibinfo  {journal} {J. Comp. Phys.}\
  }\textbf {\bibinfo {volume} {130}},\ \bibinfo {pages} {034106} (\bibinfo
  {year} {2009})}\BibitemShut {NoStop}%
\bibitem [{\citenamefont {Schaudinnus}\ \emph {et~al.}(2015)\citenamefont
  {Schaudinnus}, \citenamefont {Bastian}, \citenamefont {Hegger},\ and\
  \citenamefont {Stock}}]{Schaudinnus15}%
  \BibitemOpen
  \bibfield  {author} {\bibinfo {author} {\bibfnamefont {N.}~\bibnamefont
  {Schaudinnus}}, \bibinfo {author} {\bibfnamefont {B.}~\bibnamefont
  {Bastian}}, \bibinfo {author} {\bibfnamefont {R.}~\bibnamefont {Hegger}}, \
  and\ \bibinfo {author} {\bibfnamefont {G.}~\bibnamefont {Stock}},\
  }\href@noop {} {\bibfield  {journal} {\bibinfo  {journal} {Phys. Rev. Lett.}\
  }\textbf {\bibinfo {volume} {115}},\ \bibinfo {pages} {050602} (\bibinfo
  {year} {2015})}\BibitemShut {NoStop}%
\bibitem [{\citenamefont {Schaudinnus}\ \emph {et~al.}(2016)\citenamefont
  {Schaudinnus}, \citenamefont {Lickert}, \citenamefont {Biswas},\ and\
  \citenamefont {Stock}}]{Schaudinnus16}%
  \BibitemOpen
  \bibfield  {author} {\bibinfo {author} {\bibfnamefont {N.}~\bibnamefont
  {Schaudinnus}}, \bibinfo {author} {\bibfnamefont {B.}~\bibnamefont
  {Lickert}}, \bibinfo {author} {\bibfnamefont {M.}~\bibnamefont {Biswas}}, \
  and\ \bibinfo {author} {\bibfnamefont {G.}~\bibnamefont {Stock}},\
  }\href@noop {} {\bibfield  {journal} {\bibinfo  {journal} {J. Comp. Phys.}\
  }\textbf {\bibinfo {volume} {145}},\ \bibinfo {pages} {184114} (\bibinfo
  {year} {2016})}\BibitemShut {NoStop}%
\bibitem [{\citenamefont {Schnurr}\ \emph {et~al.}(1997)\citenamefont
  {Schnurr}, \citenamefont {Gittes}, \citenamefont {MacKintosh},\ and\
  \citenamefont {Schmidt}}]{schnurr:1997}%
  \BibitemOpen
  \bibfield  {author} {\bibinfo {author} {\bibfnamefont {B.}~\bibnamefont
  {Schnurr}}, \bibinfo {author} {\bibfnamefont {F.}~\bibnamefont {Gittes}},
  \bibinfo {author} {\bibfnamefont {F.}~\bibnamefont {MacKintosh}}, \ and\
  \bibinfo {author} {\bibfnamefont {C.}~\bibnamefont {Schmidt}},\ }\href@noop
  {} {\bibfield  {journal} {\bibinfo  {journal} {Macromolecules}\ }\textbf
  {\bibinfo {volume} {30}},\ \bibinfo {pages} {7781} (\bibinfo {year}
  {1997})}\BibitemShut {NoStop}%
\bibitem [{\citenamefont {Mokshin}\ \emph {et~al.}(2005)\citenamefont
  {Mokshin}, \citenamefont {Yulmetyev},\ and\ \citenamefont
  {H{\"a}nggi}}]{mokshin:2005}%
  \BibitemOpen
  \bibfield  {author} {\bibinfo {author} {\bibfnamefont {A.~V.}\ \bibnamefont
  {Mokshin}}, \bibinfo {author} {\bibfnamefont {R.~M.}\ \bibnamefont
  {Yulmetyev}}, \ and\ \bibinfo {author} {\bibfnamefont {P.}~\bibnamefont
  {H{\"a}nggi}},\ }\href@noop {} {\bibfield  {journal} {\bibinfo  {journal}
  {Physical review letters}\ }\textbf {\bibinfo {volume} {95}},\ \bibinfo
  {pages} {200601} (\bibinfo {year} {2005})}\BibitemShut {NoStop}%
\bibitem [{\citenamefont {Shin}\ \emph {et~al.}(2010)\citenamefont {Shin},
  \citenamefont {Kim}, \citenamefont {Talkner},\ and\ \citenamefont
  {Lee}}]{shin:2010}%
  \BibitemOpen
  \bibfield  {author} {\bibinfo {author} {\bibfnamefont {H.~K.}\ \bibnamefont
  {Shin}}, \bibinfo {author} {\bibfnamefont {C.}~\bibnamefont {Kim}}, \bibinfo
  {author} {\bibfnamefont {P.}~\bibnamefont {Talkner}}, \ and\ \bibinfo
  {author} {\bibfnamefont {E.~K.}\ \bibnamefont {Lee}},\ }\href@noop {}
  {\bibfield  {journal} {\bibinfo  {journal} {Chemical Physics}\ }\textbf
  {\bibinfo {volume} {375}},\ \bibinfo {pages} {316} (\bibinfo {year}
  {2010})}\BibitemShut {NoStop}%
\bibitem [{\citenamefont {C{\'o}rdoba}\ \emph {et~al.}(2012)\citenamefont
  {C{\'o}rdoba}, \citenamefont {Schieber},\ and\ \citenamefont
  {Indei}}]{cordoba:2012}%
  \BibitemOpen
  \bibfield  {author} {\bibinfo {author} {\bibfnamefont {A.}~\bibnamefont
  {C{\'o}rdoba}}, \bibinfo {author} {\bibfnamefont {J.~D.}\ \bibnamefont
  {Schieber}}, \ and\ \bibinfo {author} {\bibfnamefont {T.}~\bibnamefont
  {Indei}},\ }\href@noop {} {\bibfield  {journal} {\bibinfo  {journal} {Physics
  of Fluids}\ }\textbf {\bibinfo {volume} {24}},\ \bibinfo {pages} {073103}
  (\bibinfo {year} {2012})}\BibitemShut {NoStop}%
\bibitem [{\citenamefont {Darve}\ \emph {et~al.}(2009)\citenamefont {Darve},
  \citenamefont {Solomon},\ and\ \citenamefont {Kia}}]{darve:2009}%
  \BibitemOpen
  \bibfield  {author} {\bibinfo {author} {\bibfnamefont {E.}~\bibnamefont
  {Darve}}, \bibinfo {author} {\bibfnamefont {J.}~\bibnamefont {Solomon}}, \
  and\ \bibinfo {author} {\bibfnamefont {A.}~\bibnamefont {Kia}},\ }\href@noop
  {} {\bibfield  {journal} {\bibinfo  {journal} {Proceedings of the National
  Academy of Sciences}\ }\textbf {\bibinfo {volume} {106}},\ \bibinfo {pages}
  {10884} (\bibinfo {year} {2009})}\BibitemShut {NoStop}%
\bibitem [{\citenamefont {Carof}\ \emph {et~al.}(2014)\citenamefont {Carof},
  \citenamefont {Vuilleumier},\ and\ \citenamefont {Rotenberg}}]{carof:2014}%
  \BibitemOpen
  \bibfield  {author} {\bibinfo {author} {\bibfnamefont {A.}~\bibnamefont
  {Carof}}, \bibinfo {author} {\bibfnamefont {R.}~\bibnamefont {Vuilleumier}},
  \ and\ \bibinfo {author} {\bibfnamefont {B.}~\bibnamefont {Rotenberg}},\
  }\href@noop {} {\bibfield  {journal} {\bibinfo  {journal} {The Journal of
  chemical physics}\ }\textbf {\bibinfo {volume} {140}},\ \bibinfo {pages}
  {124103} (\bibinfo {year} {2014})}\BibitemShut {NoStop}%
\bibitem [{\citenamefont {Li}\ \emph {et~al.}(2017)\citenamefont {Li},
  \citenamefont {Lee}, \citenamefont {Darve},\ and\ \citenamefont
  {Karniadakis}}]{li:2017}%
  \BibitemOpen
  \bibfield  {author} {\bibinfo {author} {\bibfnamefont {Z.}~\bibnamefont
  {Li}}, \bibinfo {author} {\bibfnamefont {H.~S.}\ \bibnamefont {Lee}},
  \bibinfo {author} {\bibfnamefont {E.}~\bibnamefont {Darve}}, \ and\ \bibinfo
  {author} {\bibfnamefont {G.~E.}\ \bibnamefont {Karniadakis}},\ }\href@noop {}
  {\bibfield  {journal} {\bibinfo  {journal} {The Journal of chemical physics}\
  }\textbf {\bibinfo {volume} {146}},\ \bibinfo {pages} {014104} (\bibinfo
  {year} {2017})}\BibitemShut {NoStop}%
\bibitem [{\citenamefont {Lei}\ \emph {et~al.}(2016)\citenamefont {Lei},
  \citenamefont {Baker},\ and\ \citenamefont {Li}}]{lei:2016}%
  \BibitemOpen
  \bibfield  {author} {\bibinfo {author} {\bibfnamefont {H.}~\bibnamefont
  {Lei}}, \bibinfo {author} {\bibfnamefont {N.~A.}\ \bibnamefont {Baker}}, \
  and\ \bibinfo {author} {\bibfnamefont {X.}~\bibnamefont {Li}},\ }\href@noop
  {} {\bibfield  {journal} {\bibinfo  {journal} {Proceedings of the National
  Academy of Sciences}\ }\textbf {\bibinfo {volume} {113}},\ \bibinfo {pages}
  {14183} (\bibinfo {year} {2016})}\BibitemShut {NoStop}%
\bibitem [{\citenamefont {Ma}\ \emph {et~al.}(2016)\citenamefont {Ma},
  \citenamefont {Li},\ and\ \citenamefont {Liu}}]{ma:2016}%
  \BibitemOpen
  \bibfield  {author} {\bibinfo {author} {\bibfnamefont {L.}~\bibnamefont
  {Ma}}, \bibinfo {author} {\bibfnamefont {X.}~\bibnamefont {Li}}, \ and\
  \bibinfo {author} {\bibfnamefont {C.}~\bibnamefont {Liu}},\ }\href@noop {}
  {\bibfield  {journal} {\bibinfo  {journal} {The Journal of chemical physics}\
  }\textbf {\bibinfo {volume} {145}},\ \bibinfo {pages} {204117} (\bibinfo
  {year} {2016})}\BibitemShut {NoStop}%
\bibitem [{\citenamefont {Jung}\ \emph {et~al.}(2017)\citenamefont {Jung},
  \citenamefont {Hanke},\ and\ \citenamefont {Schmid}}]{jung:2017}%
  \BibitemOpen
  \bibfield  {author} {\bibinfo {author} {\bibfnamefont {G.}~\bibnamefont
  {Jung}}, \bibinfo {author} {\bibfnamefont {M.}~\bibnamefont {Hanke}}, \ and\
  \bibinfo {author} {\bibfnamefont {F.}~\bibnamefont {Schmid}},\ }\href@noop {}
  {\bibfield  {journal} {\bibinfo  {journal} {Journal of chemical theory and
  computation}\ }\textbf {\bibinfo {volume} {13}},\ \bibinfo {pages} {2481}
  (\bibinfo {year} {2017})}\BibitemShut {NoStop}%
\bibitem [{\citenamefont {Hansen}\ and\ \citenamefont
  {McDonald}(1990)}]{hansen:1990}%
  \BibitemOpen
  \bibfield  {author} {\bibinfo {author} {\bibfnamefont {J.-P.}\ \bibnamefont
  {Hansen}}\ and\ \bibinfo {author} {\bibfnamefont {I.~R.}\ \bibnamefont
  {McDonald}},\ }\href@noop {} {\emph {\bibinfo {title} {Theory of simple
  liquids}}}\ (\bibinfo  {publisher} {Elsevier},\ \bibinfo {year}
  {1990})\BibitemShut {NoStop}%
\bibitem [{\citenamefont {Rein~ten Wolde}\ \emph {et~al.}(1996)\citenamefont
  {Rein~ten Wolde}, \citenamefont {Ruiz-Montero},\ and\ \citenamefont
  {Frenkel}}]{tenwolde:1996}%
  \BibitemOpen
  \bibfield  {author} {\bibinfo {author} {\bibfnamefont {P.}~\bibnamefont
  {Rein~ten Wolde}}, \bibinfo {author} {\bibfnamefont {M.~J.}\ \bibnamefont
  {Ruiz-Montero}}, \ and\ \bibinfo {author} {\bibfnamefont {D.}~\bibnamefont
  {Frenkel}},\ }\href@noop {} {\bibfield  {journal} {\bibinfo  {journal} {The
  Journal of chemical physics}\ }\textbf {\bibinfo {volume} {104}},\ \bibinfo
  {pages} {9932} (\bibinfo {year} {1996})}\BibitemShut {NoStop}%
\bibitem [{\citenamefont {Steinhardt}\ \emph {et~al.}(1983)\citenamefont
  {Steinhardt}, \citenamefont {Nelson},\ and\ \citenamefont
  {Ronchetti}}]{steinhardt:1983}%
  \BibitemOpen
  \bibfield  {author} {\bibinfo {author} {\bibfnamefont {P.~J.}\ \bibnamefont
  {Steinhardt}}, \bibinfo {author} {\bibfnamefont {D.~R.}\ \bibnamefont
  {Nelson}}, \ and\ \bibinfo {author} {\bibfnamefont {M.}~\bibnamefont
  {Ronchetti}},\ }\href@noop {} {\bibfield  {journal} {\bibinfo  {journal}
  {Physical Review B}\ }\textbf {\bibinfo {volume} {28}},\ \bibinfo {pages}
  {784} (\bibinfo {year} {1983})}\BibitemShut {NoStop}%
\end{thebibliography}

\section*{Supplemental material}

\subsubsection*{Quantifying Non-Markovianity}

\begin{figure}
\begin{center}
\includegraphics[width=\linewidth]{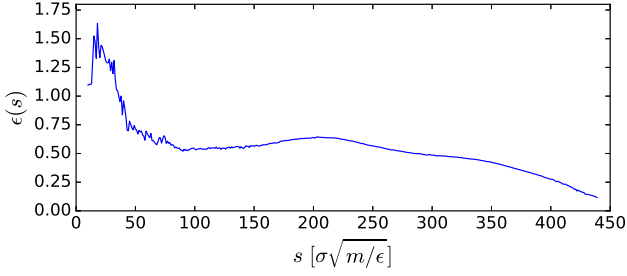}
\caption{The function $\epsilon(s)$ is clearly non-vanishing during the Lennard-Jones crystallization process. Non-Markovian features are stronger at short times than in the long-time limit.}
\label{markov_error}
\end{center}
\end{figure}

To reconstruct the full memory kernel $K(t',t)$ is an interesting way of assessing memory effects, however we can come up with a simpler measure of non-Markovianity. The Markovian limit is reached if $K(t',t) = -\gamma(t) \delta(t'-t)$, yielding $C(t',t) = C(t',t') e^{-\int_{t'}^{t} d\tau \gamma(\tau)}$ for any times $t'<t$. Therefore, for any time $s\in[t',t]$ a Markovian process fulfills $C(t',t) = C(t',s)C(s,t)$. We test the validity of this decomposition by defining a unitless quantity $\epsilon(s)$:
\begin{equation}
\label{epsilon}
    \epsilon(s) := \frac{1}{s(T-s)} \int_{0}^{s}dt' \int_{s}^{T}dt \left| 1 - \frac{C(t's)C(s,t)}{C(t',t)} \right|,
\end{equation}
where $T$ is the total time of the process. For a given value of $s$, $\epsilon(s)$ measures the deviation from the exponential decomposition at any point in the 2D-domain $\left[0,s\right]\times\left[s, T\right]$ and averages over it. By definition, $\epsilon(s)$ vanishes in the Markovian case. The dependence on $s$ allows to quantify how non-Markovian behaviour evolves throughout the process. In fig.~\ref{markov_error} we show $\epsilon(s)$ for the Lennard-Jones crystallization process, using $\tilde{C}(t',t)$ as input.

\subsubsection*{Details of the derivation for $\mathcal{J}(t',t)$}
We recall the starting point of our derivation for $\mathcal{J}$, that is 
\begin{equation}
\label{J_first_iter}
\mathcal{J}(t',t) =  j_{0}(t',t) +  \int_{t'}^{t}d\tau \mathcal{S}_{0}(t',\tau) \mathcal{J}(\tau, t) 
\end{equation}
where we have defined
\begin{align}
\label{S0}
\mathcal{S}_{0}(t',t) &= C(t',t')^{-1} \partial_{t'} C(t',t) \\
\label{j0}
j_{0}(t',t) &=  C(t',t')^{-1} \partial_{t'}\left[ C(t',t') - C(t',t) \right]
\end{align}
By substituting $\mathcal{J}(t,\tau)$ on the right-hand side of eq.~(\ref{J_first_iter}) by eq.~(\ref{J_first_iter}) itself, we obtain
\begin{align}
\label{iter_S1}
\mathcal{J}(t',t) =& j_{0}(t',t) +  \int_{t'}^{t}d\tau \mathcal{S}_{0}(t',\tau) j_{0}(\tau, t) \nonumber \\
&+ \int_{t'}^{t}d\tau \mathcal{S}_{0}(t',\tau) \int_{\tau}^{t}d\tau' \mathcal{S}_{0}(\tau,\tau') \mathcal{J}(\tau',t) \nonumber \\
=& j_{0}(t',t) +  \int_{t'}^{t}d\tau \mathcal{S}_{0}(t',\tau) j_{0}(\tau, t) \nonumber \\
&+ \int_{t'}^{t}d\tau' \left[ \int_{t'}^{\tau'} d\tau \mathcal{S}_{0}(t',\tau)  \mathcal{S}_{0}(\tau,\tau') \right] \mathcal{J}(\tau',t)\nonumber \\
=& j_{0}(t',t) +  \int_{t'}^{t}d\tau \mathcal{S}_{0}(t',\tau) j_{0}(\tau, t) \nonumber \\
&+ \int_{t'}^{t}d\tau \mathcal{S}_{1}(t',\tau) \mathcal{J}(\tau,t)
\end{align}
where we have defined $\mathcal{S}_{1}(t',t) =  \int_{t'}^{t}d\tau  \mathcal{S}_{0}(t',\tau) \mathcal{S}_{0}(\tau,t)$. We can continue and insert eqn.~(\ref{J_first_iter}) into the right-hand side of eqn.~(\ref{iter_S1}), and find
\begin{align}
\label{iter_S2}
\int_{t'}^{t}d\tau \mathcal{S}_{1}(t',\tau) \mathcal{J}(\tau,t) =& \int_{t'}^{t}d\tau \mathcal{S}_{1}(t',\tau) j_{0}(\tau,t) \nonumber \\
+ \int_{t'}^{t}d\tau &\mathcal{S}_{1}(t',\tau) \int_{\tau}^{t}d\tau' \mathcal{S}_{0}(\tau,\tau') \mathcal{J}(\tau',t) \nonumber \\
= \int_{t'}^{t}d\tau  \mathcal{S}_{1}&(t',\tau) j_{0}(\tau,t) \nonumber \\
+ \int_{t'}^{t}d\tau' &\left[\int_{t'}^{\tau'} d\tau \mathcal{S}_{1}(t',\tau)  \mathcal{S}_{0}(\tau,\tau') \right] \mathcal{J}(\tau',t)\nonumber \\
=\int_{t'}^{t}d\tau    \mathcal{S}_{1}(t',\tau) j_{0}(\tau,&t) + \int_{t'}^{t}d\tau\mathcal{S}_{2}(t',\tau) \mathcal{J}(\tau,t)
\end{align}
where we have defined $\mathcal{S}_{2}(t',t) =  \int_{t'}^{t}d\tau  \mathcal{S}_{1}(t',\tau) \mathcal{S}_{0}(\tau,t)$. By defining more generally
\begin{widetext}
\begin{equation}
\mathcal{S}_{n}(t',t) = \int_{t'}^{t} d\tau_{1} \int_{t'}^{\tau_{1}} d\tau_{2} \cdots \int_{t'}^{\tau_{n-1}}d\tau_{n} \mathcal{S}_{0}(t',\tau_{n})\mathcal{S}_{0}(\tau_{n},\tau_{n-1}) \cdots \mathcal{S}_{0}(\tau_{2}, \tau_{1})\mathcal{S}_{0}(\tau_{1},t)
\end{equation}
\end{widetext}
we can iterate the recursive substitution of $\mathcal{J}(\tau,t)$ below the integral and finally obtain
\begin{equation}
\label{J_full}
\mathcal{J}(t',t)   = j_{0}(t',t) +  \sum_{n=0}^{\infty} \int_{t'}^{t}d\tau \mathcal{S}_{n}(t',\tau) j_{0}(\tau,t)
\end{equation}

\subsubsection*{Markovian limit}

In the Markovian limit, we have $K(t',t) = \gamma(t') \delta(t'-t)$, with $\gamma(t)<0$, yielding
\begin{equation}
C(t',t) = C(t',t') \exp \left( \int_{t'}^{t} d\tau \gamma(\tau) \right)
\end{equation}
for $t'<t$. Let us compute the first orders of $\mathcal{S}_{n}(t',t)$. We first have
\begin{align}
\mathcal{S}_{0}(t',t) &= \frac{1}{C(t',t')}\frac{\partial C(t',t)}{\partial t'} = - \gamma(t')e^{\int_{t'}^{t} d\tau \gamma(\tau)} \\
\mathcal{S}_{1}(t',t) &= \int_{t'}^{t} d\tau \mathcal{S}_{0}(t',\tau)\mathcal{S}_{0}(\tau,t)  \nonumber \\
&= \gamma(t') \int_{t'}^{t}d\tau  \gamma(\tau) e^{\int_{t'}^{t} d\tau \gamma(\tau)} \nonumber \\
\mathcal{S}_{2}(t',t) &= \int_{t'}^{t} d\tau \mathcal{S}_{1}(t',\tau)\mathcal{S}_{0}(\tau,t)  \nonumber \\
&= -\gamma(t') \int_{t'}^{t} d\tau \int_{t'}^{\tau}d\tau'  \gamma(\tau') \gamma(\tau) e^{\int_{t'}^{t} d\tau \gamma(\tau)}
\end{align}
We infer
\begin{align}
\mathcal{S}_{n}(t',t) = (-1)^{n+1}&\exp\left(\int_{t'}^{t} d\tau \gamma(\tau)  \right)\gamma(t')\nonumber \\
 \times \int_{t'}^{t}&d\tau_{1} \gamma (\tau_{1}) \cdots \int_{t'}^{\tau_{n-1}}d\tau_{n} \gamma(\tau_{n}) 
\end{align}
The nested integral on the right-hand side, together with the $(-1)^{n}$ factor, consist in the $n$-th order term in the Taylor expansion of the the exponential function $\exp\left(- \int_{t'}^{t} d\tau  \gamma(\tau)  \right)$ :
\begin{align}
\sum_{n=0}^{\infty} (-1)^{n}\int_{t'}^{t}d\tau_{1} &\gamma (\tau_{1}) \cdots \int_{t'}^{\tau_{n-1}}d\tau_{n} \gamma(\tau_{n}) \nonumber \\
 &= \exp\left(- \int_{t'}^{t} d\tau  \gamma(\tau)  \right) \nonumber \\
\Rightarrow \sum_{n=0}^{\infty} \mathcal{S}_{n}(t',t)& =  -\gamma(t')\exp\left(- \int_{t'}^{t} d\tau  \gamma(\tau)  \right) \nonumber \\
&\ \ \ \ \ \ \ \ \ \ \ \ \times \exp\left( \int_{t'}^{t} d\tau  \gamma(\tau)  \right)  \nonumber \\
\Rightarrow \mathcal{S}(t',t)& =  -\gamma(t')
\end{align}
Again, this is valid for $t'<t$. In the case $t'>t$, the same calculation would yield $\mathcal{S}(t',t) = \gamma(t')$. In other words, for any $t$ and $t'$, we would have $\mathcal{S}(t',t) = -\mathcal{J}(t',t) = \gamma(t') \left[2\Theta(t'-t) - 1 \right]$, where $\Theta$ is the Heaviside function. The derivative with respect to $t$ would then yield $K(t',t) = \gamma(t')\delta(t'-t)$, which is consistent with what we started from. 

An interesting property of the functions $\mathcal{S}_{n}(t',t)$ is the fact that their derivatives with respect to $t$ can be written as
\begin{equation}
\frac{\partial \mathcal{S}_{n}(t',t)}{\partial t} = \gamma(t) \left( \mathcal{S}_{n}(t',t) - \mathcal{S}_{n-1}(t',t) \right)
\end{equation}
In other words, we have $\partial_{t} \mathcal{S}_{n}(t',t) = 0 \Leftrightarrow \mathcal{S}_{n}(t',t) = \mathcal{S}_{n-1}(t',t)$: the extremum of the function $\mathcal{S}_{n}(t',t)$ is located where it crosses the previous order. This is a graphical check that can easily be performed to test the Markovian assumption. 

\paragraph*{Stationary case}
In the stationary limit, $\gamma(t)$ becomes a constant. We then obtain, for $t'<t$
\begin{equation}
\mathcal{S}_{n}(t',t) = \frac{(-\gamma)^{n+1}(t-t')^{n}e^{\gamma(t-t')}}{n!}
\end{equation}
At constant $t'$, $\mathcal{S}_{n}(t',t)$ presents an extremum at $t^{*}_{n}=t' - n/\gamma$, and $\mathcal{S}_{n}(t',t_{n}^{*}) = -\gamma n^{n}e^{n}/n! \stackrel{n\to\infty}{\propto} n^{-1/2}$. In a stationary process, one can hence test the Markovian assumption by checking if the extrema of the functions $\mathcal{S}_{n}(t',t)$ are located with an equal spacing and if their amplitude decay as $1/\sqrt{n}$.

\end{document}